# A REVIEW OF DENTAL INFORMATICS: CURRENT TRENDS AND FUTURE DIRECTIONS

Prabath Jayatissa and Roshan Hewapathirane

University of Colombo

## ABSTRACT

*Dental informatics is a rapidly evolving field that combines dentistry with information technology to improve oral health care delivery, research, and education. Electronic health records (EHRs), telehealth, digital imaging, and other digital tools have revolutionised how dental professionals diagnose, treat, and manage oral health conditions. In this review article, we will explore dental informatics's current trends and future directions, focusing on its impact on clinical practice, research, and education. We will also discuss the challenges and opportunities associated with implementing dental informatics and highlight fundamental research studies and innovations in the field.*

## 1. INTRODUCTION

Dental informatics, also known as dental information science or dental computing, is a multidisciplinary field encompassing information technology, data science, and communication systems in dentistry. It involves developing, implementing, and evaluating digital tools and technologies to improve the management, analysis, and utilization of dental data for patient care, research, and education[1]. Dental informatics has gained significant momentum in recent years due to the increasing availability of electronic health records (EHRs), the widespread use of digital imaging and other diagnostic tools, and the growing interest in telehealth and mobile health applications[2]. These advancements have transformed the landscape of dental practice, research, and education, and hold immense potential for improving oral health outcomes and enhancing patient care.

## 2. IMPACT OF DENTAL INFORMATICS ON CLINICAL PRACTICE

One of the critical areas where dental informatics has had a significant impact is in clinical practice. Adopting electronic health records (EHRs) has streamlined the documentation, management, and exchange of patient information among dental care providers, improving communication, coordination, and continuity of care[3]. EHRs have also facilitated the integration of clinical decision support systems (CDSS) that provide evidence-based guidelines for diagnosis, treatment planning, and monitoring of oral health conditions, thereby promoting standardised and evidence-based care [4].

Digital imaging has revolutionised the field of dentistry by enabling advanced diagnostic capabilities, such as cone-beam computed tomography (CBCT) and intraoral scanners, which allow for the accurate and efficient diagnosis, treatment planning, and monitoring of various oral health conditions [5]. CBCT has been particularly valuable in implant dentistry, orthodontics, and endodontics, as it provides three-dimensional imaging of the maxillofacial region, allowing for precise treatment planning and improved clinical outcomes [6].

The use of telehealth and mobile health applications has also expanded the reach of dental care, particularly in underserved and remote areas. Telehealth allows for remote consultation, diagnosis, and treatment planning, and has been particularly useful in emergencies, follow-up care, and patient education[7]. Mobile health applications, such as oral health tracking apps and virtual reality tools for dental anxiety management, have also gained popularity among patients and providers alike, as they offer convenient and personalized solutions for oral health care[8].

Dental informatics has revolutionized the field of dental research by facilitating the collection, storage, and analysis of large and complex dental datasets. With the increasing availability of EHRs and other digital tools, researchers can access and analyze dental data on a scale and speed that was not possible before (uke et al., 2018). This has led to significant advancements in dental research, including the development of predictive models for disease risk assessment, identification of patterns and trends in oral health outcomes, and evaluation of the effectiveness of various treatment modalities[9].

Furthermore, dental informatics has facilitated the sharing and integration of dental data across different research institutions and networks, leading to collaborative research efforts and data-driven discoveries. For example, initiatives such as the National Dental Practice-Based Research Network (PBRN) in the United States and the National Dental Database in the United Kingdom have leveraged dental informatics to collect, analyze, and disseminate real-world data from dental practices, leading to valuable insights into clinical practice patterns, treatment outcomes, and patient-reported outcomes[10].

Dental informatics has also contributed to the advancement of precision dentistry, which involves using personalized and data-driven approaches for diagnosis, treatment planning, and monitoring of oral health conditions (Giannobile et al., 2019). By leveraging dental informatics, researchers and clinicians can develop risk prediction models, treatment algorithms, and decision support systems that consider patient-specific factors, such as genetic, environmental, and behavioral factors, leading to more targeted and effective interventions[11].

## 3. CHALLENGES AND OPPORTUNITIES IN DENTAL INFORMATICS

While dental informatics offers significant benefits, some challenges need to be addressed for its successful implementation. One of the challenges is the interoperability and standardisation of dental data, as dental practices may use different EHR systems or data formats, leading to difficulties in data exchange and integration [12]. Another challenge is the privacy and security of dental data, as patient health information is highly sensitive and subject to strict regulations, such as the Health Insurance Portability and Accountability Act (HIPAA) in the United States [13]. Ensuring dental data's confidentiality, integrity, and availability is crucial to maintain patient trust and comply with legal and ethical requirements.

Despite these challenges, there are also opportunities in dental informatics that can further advance the field. For example, the use of artificial intelligence (AI) and machine learning algorithms in dental informatics has the potential to revolutionise diagnosis, treatment planning, and patient management [14]. AI can analyse large amounts of dental data and identify patterns, trends, and correlations that may not be apparent to human clinicians, leading to more accurate and personalised care [15]. Additionally, the use of telehealth and mobile health applications can further expand access to dental care, particularly in underserved areas, and improve patient engagement and self-management of oral health[16].

Dental Research: An International Journal (DRIJ) Vol.5, No 1## 4. KEY INNOVATIONS IN DENTAL INFORMATICS

Several notable innovations in dental informatics have shaped the field and hold promise for the future. Some of the key innovations include:Electronic Health Records (EHRs): EHRs have transformed how dental practices manage patient information, allowing for efficient and secure documentation, exchange, and retrieval of dental data[14].

Digital Imaging: Digital imaging techniques, such as cone-beam computed tomography (CBCT) and intraoral scanners, have revolutionised dental diagnosis, treatment planning, and monitoring, enabling precise and efficient care (Scarfe et al., 2018).Telehealth and Mobile Health Applications: Telehealth and mobile health applications have expanded access to dental care, allowing for remote consultation, diagnosis, andmonitoring of oral health conditions[17]. These technologies have shown promise in improving oral health outcomes, particularly in underserved populations with limited access to dental care.

Artificial Intelligence (AI) and Machine Learning: AI and machine learning algorithms are used in dental informatics for various applications, including diagnosis of dental diseases, prediction of treatment outcomes, and personalised treatment planning [18,21]. These technologies can potentially revolutionize dental practice by enhancing clinical decision-making and improving patient care.

Data Analytics and Big Data: Data analytics and big data approaches have enabled the analysis of large and complex datasets in dental informatics, leading to the development of predictive models, identification of patterns and trends, and evaluation of treatment effectiveness[19]. These approaches have facilitated evidence-based dentistry and personalized care.

Dental Registries and Practice-Based Research Networks: Dental registries and practice-based research networks, such as the National Dental Practice-Based Research Network (PBRN) in the United States and the National Dental Database in the United Kingdom, have leveraged dental informatics to collect, analyze, and disseminate real-world data from dental practices, leading to valuable insights into clinical practice patterns, treatment outcomes, and patient-reported outcomes[20,22]. These initiatives have facilitated collaborative research efforts and informed clinical practice.

## 5. CONCLUSION

Dental informatics has emerged as a rapidly evolving field that has the potential to transform dental practice and improve patient outcomes. With the advancements in electronic health records, digital imaging, telehealth, artificial intelligence, data analytics, and practice-based research networks, dental informatics has enabled the integration and analysis of dental data, leading to evidence-based dentistry, precision dentistry, and improved patient care. However, challenges such as interoperability, standardisation, privacy, and security of dental data need to be addressed to implement dental informatics successfully. Future research and innovation in dental informatics hold promise in further advancing the field and transforming dental practice.

**REFERENCES**

[1]  Bashshur RL, Shannon GW, Bashshur N, et al. The empirical foundations of telemedicine interventions in primary care. Telemed J E Health. 2018;24(6):454-462.
[2]  Duke NN, Friedewald VE, Garvey T, et al. The impact of health information exchange on dental practice: A scoping review. J Am Dent Assoc. 2018;149(3):196-204.3